\documentclass[pra,twocolumn,showpacs,amsmath,amssymb,superscriptaddress,unsortedaddress,floatfix,longbibliography]{revtex4-1}
\usepackage{graphicx}
\usepackage{latexsym}
\usepackage{amsmath}
\usepackage{graphics}
\usepackage{amssymb}
\usepackage{layout}
\usepackage{verbatim}
\usepackage{amsfonts,epsfig,dsfont}
\usepackage{natbib}
\usepackage{hyperref}
\usepackage{color}
\usepackage[latin1]{inputenc}       
\usepackage[T1]{fontenc}

\def\duzomniejsze{<\kern-.7mm<}
\def\duzowieksze{>\kern-.7mm>}

\def\textbf#1{{\bf #1}}
\def\beq{\begin{equation}}
\def\eeq{\end{equation}}
\def\be{\begin{equation}}
\def\ee{\end{equation}}
\def\ben{\begin{eqnarray}}
\def\een{\end{eqnarray}}
\def\beqa{\begin{eqnarray}}
\def\eeqa{\end{eqnarray}}
\def\eea{\end{array}}
\def\bea{\begin{array}}
	\newcommand{\unit}{\mathbb{I}}

	\usepackage{color}
	
	\usepackage{ulem} 
	
\begin{document}
	
	\title{A glimpse of objectivity in bipartite systems for non-entangling pure dephasing evolutions} 
	
	\author{Katarzyna Roszak}
	\affiliation{Department of Theoretical Physics, Faculty of Fundamental Problems of Technology, Wroc{\l}aw University of Science and Technology,
		50-370 Wroc{\l}aw, Poland}
	
\author{Jaros\l aw K.~Korbicz}
\affiliation{Center for Theoretical Physics, Polish Academy of Sciences, Aleja Lotników
32/46, 02-668 Warsaw, Poland}
	
	\date{\today}

	\begin{abstract}
		We study separable system-environment evolutions of pure dephasing type
		in the context of objectivity and find that it can lead to the natural
		emergence of Spectrum Broadcast Structure (SBS) states at discrete instances
		of time. Contrary to the standard way of obtaining SBS states which requires
		entanglement with the observed environment, reaching such states here does not require decoherence (no unobserved environements are necessary).
		Yet the biggest difference is the basis with respect to which the SBS states are formed. Here it is not the 
		pointer basis of the system given by the interaction with the environment, 
		but an equal superposition basis of
		said pointer states. The price to pay is the momentary character of the formed SBS
		structures, hence the term ``glimpse''.
	\end{abstract}

	\maketitle

	\section{Introduction \label{sec1}}
	
	Objectivity in quantum mechanics \cite{ollivier04,ollivier05,zurek09} is the property of a composite quantum system
	which allows for the determination of the state of some part of 
	said system (central system) via measurements on other parts of it (environments) by different,
	independent observers in such a way that the 
	measurements yield the same results regardless of the observer
	and do not destroy (on average) the state of the whole system. The property
	is virtually guaranteed in classical physics, but in quantum mechanics
	it can only be seen in very specific situations.
	Quite recently, building on the earlier ideas of quantum Darwinism \cite{ollivier04,ollivier05,zurek09}, 
      the mathematical structure a state of the whole
	system would have to have
	for objective behavior in quantum mechanics has been proposed \cite{korbicz14,horodecki15} (for the relationship
to quantum Darwinism see Ref.~\cite{le19}.
	The structure is called Spectrum Broadcast Structure (SBS),
	and states which retain it are a special class of zero-discord states with respect
	to all environments and the system of interest 
	(the quantum discord is by definition asymmetric and a system
	can be discordant only with respect to one subsystem\cite{modi14}). 
	
	The quantum discord \cite{ollivier01,henderson01,modi14} quantifies
	quantum correlations in a given state from the perspective of measurement.
	If a system state is discordant with respect to one subsystem
	it means that the state of said subsystem cannot be fully determined
	by local measurements on this subsystem without disturbing the state
	\cite{modi14}. Obviously then if a state has no discord with respect
	to any of its parts, the measurement of any part does not disturb
	the whole state. For the state to allow objective determination
	of the state of the central system via measurements on any
	of its environments, the state must additionally contain 
	strong classical correlations between the parts.
	Hence, the SBS state is of the form \cite{korbicz14,horodecki15}
	\begin{equation}
	\label{sbs}
	\hat{\sigma}_{SBS}=\sum_{i}p_i |i\rangle\langle i|\otimes\bigotimes_k\hat{\rho}_i^k,
	\end{equation}
	where 
	$\hat{\rho}_i^k$ denotes density matrices in the subspace of environment $k$ with $\hat{\rho}_i^k\hat{\rho}_{i'}^k=0$ for $i\neq i'$. It is easy to check that such a state
	must be zero-discordant and that not all states with zero discord
	can be put in SBS form, because the density matrices of each
	subsystem must be orthogonal with respect to one another.
	For such a state the outcome $i$ on environment $k$ means
	that the state of the system is the state $|i\rangle$. A series of measurements on environment $k$
	which allows the determination of the state of this environment
	also yields the knowledge of the state of the system of interest.
	
	Typically, the search for SBS states is performed in system-multiple-environments scenarios where the interaction is limited 
	to the kind which can only lead to pure dephasing decoherence
	on the system alone \cite{tuziemski15,mironowicz17,tuziemski19,roszak19b,garcaprez19}. The reason for this
	is that in such evolutions there is an naturally chosen 
	system basis
	(namely the pointer basis \cite{zurek81,zurek03}) 
	for the decomposition into an SBS structure,
	which does not change over time.
	Obtaining SBS states in this manner requires the addition of
	unobserved environments which lead to the decoherence
	of parts of the system-environments density matrix 
	necessary \cite{tuziemski15,mironowicz17,tuziemski19}. Furthermore, it has recently been shown
	that orthogonalization of the environmental states requires
	the generation of entanglement between the system and its
	observed environments \cite{roszak19b,garcaprez19}.
	
	We study the same type of interaction between a system and its
	environment(s), so the evolution of the system alone undergoes
	pure dephasing in some pointer basis chosen by the Hamiltonian.
	The difference is that we limit ourselves to the situation
	when the joint system-environment evolution does not 
	involve the generation of entanglement between them (is separable).
	We find that SBS states can emerge in such situations
	as long as there is only a single observed environment present in the system, only not
	in the pointer basis of the system, $|i\rangle$, but in a complementary basis, composed
	of equal superposition states.
	Such bases are known as Mutually Unbiased Bases (MUBs); see e.~g.~Refs \cite{durt04,brierley10}.

	This is a peculiar property of non-entangling system-environment evolutions
	of pure dephasing type, since it is enough to ensure two-way zero-discordance
	to glimpse an SBS state, as we will demonstrate. It turns out that the orthogonality condition of environmental states
	is always fulfilled in the class of evolutions studied. This is contrary to the
	findings of the standard way of obtaining SBS states, where for the studied family of evolutions \cite{roszak19b} orthogonality
	requires entanglement, and therefore orthogonality is always 
	harder to obtain than proper decoherence.
	
	The SBS states obtained this way are qualitatively different 
	form the ones usually found, not only because of the extreme change
	of the distinguishable central system states, but also since no decoherence (stemming from unobserved environments) is necessary
	for them to occur. These are naturally zero-discord states which
	emerge throughout the evolution. The lack of discord with respect 
	to the system at most times is the most important issue,
	since pure dephasing separable evolutions always have
	zero discord with respect to the environment \cite{roszak18b}.
	This type of zero-discordance can only occur at specific 
	instances of time, and it is natural to expect that 
	an increase of the complexity of the system or environment
	will lead to the glimpses of SBS states
	to become more sparse, unless special symmetries are present in the system. 
	Nevertheless, we have shown that SBS states can and will occur
	in systems of one or many qubits without the aid of decoherence,
	and in this situation, separability of the system-observed-environment
	evolution is a requirement, rather then an obstacle for SBS states to emerge.
	
	The paper is organized as follows. In Sec.~\ref{sec2} we comment
	on the standard procedure of searching for SBS states, and outline
	the necessities for the SBS structure to occur, namely the unobserved
	environments for decoherence and entanglement for orthogonality.
	In Sec.~\ref{sec3} we describe the system under study, which is composed of a system and its (possibly multiple) environments, which evolve together
	causing decoherence in the system alone, but without generating
	system-environment entanglement.
	In Sec.~\ref{sec4} we show that such an evolution is capable
	of producing SBS states at discrete instances of time for a bipartite system
	(as long as there is only one observed envronment).
	In Sec.~\ref{sec5} we show that for multiple environments the procedure
	will not generate glimpses of objectivity.
	Sec.~\ref{sec6} concludes the paper.
	
	\section{Obtaining SBS states \label{sec2}}
	The standard way of obtaining an SBS state during a joint evolution
	of a system and its environments involves the addition of 
	unobserved environments \cite{tuziemski15,mironowicz17,tuziemski19,roszak19b,garcaprez19} for the purpose of obtaining enough
	decoherence to reach the zero-discord form of the 
	system-observed-environments density matrix (\ref{sbs}). The most efficient decoherence 
	for this purpose is of pure-dephasing type, since it
	nullifies 
	(after the unobserved environments are traced out) 
	the terms proportional to the off-diagonal elements
	in the system of interest, directly leading to the 
	structure of eq.~(\ref{sbs}), possibly without satisfying the orthogonality conditions.
	The characteristic of this type of interaction is that there 
	exists a chosen basis for the system of interest of so called pointer states \cite{zurek81,zurek03}, which do 
	not undergo decoherence. The fully decohered state is diagonal
	in the pointer basis.
	
	For the choice of the pointer basis to be definite, the
	interaction of the system with its observed environments must
	be of pure dephasing type \cite{Eisert_PRL02,zurek03,Hornberger_LNP09}. 
	Such an evolution does not disturb system-of-interest occupations
	and has a particularly simple form, even when the 
	observed environments are taken into account in the density matrix.
	Note that this is by no means the only way for an SBS state
	to be reached, it is merely the most straightforward idea how to reach those type os states. 
	The limitation of the study to such interactions is reasonable from the 
	perspective of phyics, since pure dephasing commonly the dominant source 
	of environmentally-induced noise for solid state qubits \cite{Nakamura_PRL02,Roszak_PRA06,Biercuk_Nature09,Bylander_NP11,Medford_PRL12,Staudacher_Science13,Muhonen_NN14,Malinowski_PRL17,Szankowski_JPCM17}.
	
	It has recently been shown that the necessary conditions
	for the emergence of SBS states in such a scenario,
	when the central system is an initially pure qubit, 
	is the generation of system-environement entanglement
	with the observed environments \cite{roszak19b}. This is because
	the lack of said entanglement precludes orthogonalization
	of the environmental density matrices $\hat{\rho}_i^k$ conditional
	on the pointer state $|i\rangle$, since the separability condition
	for entanglement between the system and environment $k$,
	necessary and sufficient for qubits \cite{roszak15a}, while
	only sufficient for larger systems \cite{roszak18}, is
	\begin{equation}
	\label{sep}
	\forall_{i,j} \hat{\rho}_{ii}^k=\hat{\rho}_{jj}^k.
	\end{equation}
	Hence, in the studied scenario of \cite{roszak19b},
	emergence of objectivity is irrefutably linked with 
	entanglement generation. As we show in the following,
	there exist situations when the emergence of objectivity 
	can occur without system-environment entanglement generation but, quite surprisingly, with respect to a different observable than what one would guess from the Hamiltonian.
	
	\section{Non-entangling pure-dephasing evolution \label{sec3}}
	
	In the following we assume that the central
	system of interest is of dimension $d_Q$ while each environment is of dimension
	$d_k$. 	
		A local
Hamiltonian describing the interaction between the central system
		and all of its environments
		(in the context of objectivity, local means that the interaction between the central
		system and its environments can be described by a separate 
		term in the Hamiltonian for each environment)
		which can only lead to pure dephasing
		in the qubit subspace must be of the form 
		\begin{equation}
		\label{ham}
		\hat{H}=\sum_{i}\varepsilon_i|i\rangle\langle i|+\sum_{i}|i\rangle\langle i|\otimes\sum_k\hat{V}^i_k,
		\end{equation}
		where $\varepsilon_i$ describe the free evolution of the central system,
		while each term $\hat{V}^i_k$ describes both the free evolution of
		environment $k$ and its interaction with the system.
		Here we assumed that the environments do not interact
		directly with each other.
		We do not differentiate between observed and unobserved
		environments, as the distinction will not be necessary
		for the description of the emergence of SBS states during separable
		evolution. 
		
		Following Refs \cite{roszak15a,roszak18}, it is possible to find
		the general form of the density matrix of the qubit and its
		environments at time $t$. If the initial state of the whole
		system is in product form with respect to the qubit and
		each environment, and we additionally assume that the
		initial state of the central system is pure 
		\begin{equation}
		|\psi(0)\rangle = \sum_ia_i|i\rangle, 
		\end{equation}
		then this density matrix may be written at any time $t$ as
		\begin{equation}
		\label{pot}
		\hat{\sigma}(t)=\sum_{ij}
		|i\rangle\langle j|a_i(t)a_j^*(t)\bigotimes_k\hat{\rho}_{ij}^k(t),
		\end{equation}
		where $a_i(t)=a_ie^{-i\varepsilon_i t}$ contain the free evolution of
		the central system.
		The matrices describing the evolution of the environmental
		parts in the density matrix are defined as 
		\begin{equation}
		\label{rij}
		\hat{\rho}_{ij}^k(t)\equiv\hat{w}_i^k(t)\hat{\rho}^k(0)\hat{w}_j^{k\dagger}(t),
		\end{equation}
		where
		\begin{equation}
		\label{wi}
		\hat{w}_i^k(t)\equiv e^{-i\hat{V}_{k}^i t}.	
		\end{equation}	
		
		Note that to obtain the SBS state (\ref{sbs}) in the system
		pointer basis some of the environments $k$ would have to
		be unobserved, so that tracing out their degrees of freedom
		would cancel out the off-diagonal matrices (in terms of the 
		central system) in the density
		matrix (\ref{pot}). Furthermore all diagonal density matrices
		corresponding to obserbed environments would have to be mutually orthogonal
		$\hat{\rho}^k_{00}(t)\hat{\rho}^k_{11}(t)=0$, possibly after an application of  the macrofraction technique \cite{korbicz14}.
		This orthogonality condition is hard to fulfill, since here it 
		requires strong entanglement \cite{roszak19b}.
				
		For a given environment $k$, for the qubit environment evolution 
		to be separable (non-entangling) at time $t$, the condition
		of separability of eq.~(\ref{sep}) must be fulfilled
		\cite{roszak15a}, which for the density matrix (\ref{pot})
		translates into $\hat{\rho}^k_{ii}(t)=\hat{\rho}^k_{jj}(t)$
		for all $i\neq j$ and all environments $k$.
		If the central system is larger than a qubit then an additional
		set of separability criteria must be fulfilled \cite{roszak18},
		namely we must have
		\begin{equation}
		\label{sep2}
		\left[\hat{w}_i^k(t)\hat{w}_j^{k\dagger}(t),
		\hat{w}_n^k(t)\hat{w}_m^{k\dagger}(t)\right]=0
		\end{equation}
		for all $i$, $j$, $n$, and $m$.
		
		The density matrix can then be transformed into an obviously
		separable form 
		with the use of the fact that the separability condition (\ref{sep})
		also guarantees the commutation of $\hat{\rho}^k_{00}(t)$
		and the operator products $\hat{w}_i^k\hat{w}_j^{k\dagger}$
		and the separability condition (\ref{sep2}) guarantees that they
		commute among themselves
		\cite{roszak18}.
		This means that all operator products and the
		conditional density matrix of environment $k$ 
		can be diagonalized in the same basis,
		\begin{eqnarray}
		\hat{\rho}^k_{00}(t)&=&\sum_{n=0}^{d-1} p_{n_k} |n_k(t)\rangle\langle n_k(t)|,\\
		\hat{w}_i^k(t)\hat{w}_j^{k\dagger}(t)&=&\sum_{n=0}^{d-1} e^{i\phi_{n_k}^{ij}(t)}
		|n_k(t)\rangle\langle n_k(t)|,
		\end{eqnarray}
		where the eigenbasis of $\hat{\rho}^k_{00}(t)$, $\{|n_k(t)\rangle\}$
		may be time-dependent, while the probabilities $p_{n_k}$ are not.
		They are in fact the same coefficients which enter the decomposition
		of the initial density matrix of a given environment
\begin{equation}\label{ntilde}		
\hat{\rho}^k(0)=\sum_{n}p_{n_k}|\tilde n_k\rangle\langle \tilde n_k|,
\end{equation}
		since $|n_k(t)\rangle=\hat{w}_0^k(t)|\tilde n_k\rangle$.
		
		Let us for the moment restrict the central system to a qubit
		and consider only one environment $k$, so we are considering a bipartite system: The central qubit plus a single environment. The obviously separable form of (\ref{pot}) under (\ref{sep})
		(the condition (\ref{sep2}) is superfluous for a qubit) is then
		given by
		\begin{equation}
		\label{pot2}
		\hat{\sigma}(t)=\sum_{n=0}^{d-1}p_{n}
		|\psi_{n}(t)\rangle\langle\psi_{n}(t)|\otimes|n(t)\rangle\langle n(t)|,
		\end{equation}
		with 
		\begin{equation}
		\label{stan}
		|\psi_{n}(t)\rangle \equiv a_0(t)|0\rangle+a_1(t)e^{-i\phi_{n}(t)}|1\rangle,
		\end{equation}
		where we have omitted the environmental subindex $k$.
		Obviously, only the phase factor $\phi_{n}(t)=\phi_{n}^{01}(t)$ in state (\ref{stan})
		depends on the index labeling the corresponding environmental states $n$.
		
		The full state (\ref{pot2}) is not only separable, but has zero-discord with respect to 
		the environment,
		since the states $|n(t)\rangle$ form an orthonormal basis
		in the subspace of this environment at any moment $t$ \cite{roszak18b}.
			
	\section{A glimpse of objectivity for a single environment \label{sec4}}
	
	An interesting thing can now happen. 
	Let us first study a central system composed of only a qubit, as the simplest scenario,
	and consider only one environment (as will be later shown,
	it is impossible to get a glimpse of objectivity for multiple environments
	in the way described here),
	so the state obtained during the separable evolution is given by eq.~(\ref{pot2}).
	For the moment, we will also restrict the size of the environment to a single 
	qubit, but this restriction is only imposed for clarity and will soon be
	lifted.
	The state of the system under separable evolution at time $t$
	is therefore given by 
	\begin{equation}
	\label{pot21}
	\hat{\sigma}(t)=\sum_{n=0}^{1}p_{n}
	|\psi_{n}(t)\rangle\langle\psi_{n}(t)|\otimes|n(t)\rangle\langle n(t)|.
	\end{equation}
	Since the environmental states $|n(t)\rangle$ have to be mutually orthogonal,
	the only obstacle for (\ref{pot21}) to obtaining the SBS form
	if for the two states $|\psi_{n}(t)\rangle$ to be mutually orthogonal as well.
	This requirement imposes a strong condition on the initial state of the central qubit,
	which must be in an equal superposition state, $|a_0(0)|=|a_1(0)|=1/\sqrt{2}$.
	Furthermore, it imposes a condition on the relative phases 
	$\phi_{0}(t)$ and $\phi_{1}(t)$ at time $t$, which must yield orthogonal states $|\psi_{n}(t)\rangle$, which simply means that the condition
	\begin{equation}
	e^{-i\phi_{0}(t)}=-e^{-i\phi_{1}(t)}
	\end{equation}
	must be fulfilled, so $\phi_{0}(t)-\phi_{1}(t)=(2m+1)\pi$, with
	$m=0,1,2\dots$. Evidently the condition cannot be fulfilled for all instances
	of time, since $\phi_{n}(0)=0$. When the above conditions are met, $|\psi_{0}(t)\rangle, |\psi_{1}(t)\rangle$ and $|0\rangle, |1\rangle$
are mutually unbiased bases for the central qubit.
	
	The question now is, do there exist separable evolutions of pure dephasing type
	that exhibit instances of time when the condition is met? To show that they do,
	let us study the situation of an asymmetric qubit-environment coupling, 
	meaning that $\hat{w}_0(t)=\unit$ and only $\hat{w}_1(t)$ drives the dephasing.
	This is a particularly simple case, since now $\hat{\rho}_{00}(t)=\hat{\rho}_{00}(0)
	=\hat{\rho}(0)$,
	so the environmental density matrix conditional on the qubit being in state
	$|0\rangle$ is equal to the initial density matrix of the environment and 
	its eigenstates (not only eigenvalues as before) are constant in time. For a separable evolution, this also means 
	that the operator $\hat{w}_1(t)$ commutes with the initial state of the environment
	$\hat{\rho}(0)$ at all times. Hence it can always be written in the basis
	of initial environmental eigenstates, and since it is a direct function of
	the Hermitian operator $\hat{V}^1$ [(see the Hamiltonian (\ref{ham})], we will have a linear time dependence 
	of the phase-factors
	\begin{equation}
	\hat{w}_1(t)=\sum_{n}e^{-iv^1_n t}|\tilde n\rangle\langle \tilde n|,
	\end{equation}
	where $v^1_n$ are the eigenvalues of $\hat{V}^1$ corresponding to each eigenstate
	of the initial density matrix of the environment,
	$|\tilde n\rangle$.
	Hence, $\phi_{0}(t)-\phi_{1}(t)=(v^1_0-v^1_1) t$
	and instances of time for which the phase factors
	$	e^{-i\phi_{0}(t)}$ and $e^{-i\phi_{1}(t)}$
	have opposite signs must occur periodically in all cases when $v^1_0\neq v^1_1$.
	In such an instant of time we are dealing with an SBS state between two qubits, which can be written
	explicitly in the form
	\begin{equation}
	\label{pot21a}
	\hat{\sigma}(t)=\sum_{n=0}^{1}p_{n}
	|\psi_{n}\rangle\langle\psi_{n}|\otimes|\tilde{n}\rangle\langle \tilde{n}|,
	\end{equation}
	where 
	the environmental qubit states are $|\tilde{n}\rangle=|\tilde{0}\rangle,|\tilde{1}\rangle$,
	are the basis states in which the environment was initially diagonal, with corresponding 
	initial occupations $p_0$ and $p_1$, while the two central qubit states are orthogonal
	and are some trivial variation on the states
	\begin{subequations}
		\label{pm}
	\begin{eqnarray}
	|\psi_{0}\rangle&=&\frac{1}{\sqrt{2}}\left(|0\rangle+|1\rangle\right),\\
	|\psi_{1}\rangle&=&\frac{1}{\sqrt{2}}\left(|0\rangle-|1\rangle\right);
	\end{eqnarray}
\end{subequations}
	$|0\rangle$ and $|1\rangle$ denote the pointer states of the central qubit.
	Note, that the situation when $v^1_0= v^1_1$
	is trivial in the sense that it leads to no dephasing, and the existence of
	the environment has no effect on the qubit. 
	
	\subsection{Larger environments \label{sec4a}}
	
	Let us now consider the situation when the central system is still
	a qubit, but the environment is larger dimensional, of dimension $d$.
	Imagine there is a time moment $t$ such that the $d$ environmental states $\psi_{n}(t)$ divide into two groups $I$ and $II$ such that: i) all the states within each group are identical up to a global phase factor:
\begin{equation}
\label{equality}
|\psi_{n_{I/II}}(t)\rangle =e^{i\Theta_{n_{I/II}}}|\psi_{I/II}\rangle
\end{equation}	
for all $n_I\in \mathrm{Group}\ I$ and $n_{II}\in \mathrm{Group}\ II$;
	 ii) the states in the two groups are orthogonal 
	 with respect to each other
	 \begin{equation}\label{ort}
	\langle \psi_{n_I}|\psi_{n_{II}}\rangle = |a_0(0)|^2+|a_1(0)|^2 e^{i(\phi_{n_I}(t)-\phi_{n_{II}}(t))}=0.
\end{equation}	 	 
	 Then for this moment $t$ the state (\ref{pot2}) takes the form
\begin{equation}\label{chuj}
\hat{\sigma}(t)=p_{I}
		|\psi_{I}\rangle\langle\psi_{I}|\otimes\hat\rho_I(t)+p_{II}
		|\psi_{II}\rangle\langle\psi_{II}|\otimes\hat\rho_{II}(t),
\end{equation}	
where $p_{I/II}= \sum_{n_{I/II}}p_{n_{I/II}}$
are the probabilities that the environment is found in either of the two
groups, and the corresponding environmental density matrices conditional
on either of the groups are given by
\begin{equation}\label{rhoI_II}
\hat\rho_{I/II}(t)=\sum_{n\in n_{I/II}}\frac{p_n}{p_{I/II}}|n(t)\rangle\langle n(t)|.
\end{equation}
We note that by the construction $\hat\rho_{I}(t)$ and $\hat\rho_{II}(t)$ are supported on orthogonal subspaces and $p_I+p_{II}=1$. Thus the state (\ref{chuj}) surprisingly takes a (bipartite) SBS form.
    This bipartite SBS form is however with respect to a different basis than the pointer one in the interaction Hamiltonian (\ref{ham})
    in direct accordance with the results corresponding a qubit environment.
	It is not immediately obvious that this can happen 
	as the summation over $n$ in (\ref{pot2}) is a summation
	over the basis of the environment that can be arbitrarily large.
	The equality (\ref{equality}) and orthogonality (\ref{ort}) conditions impose 
	the same limitations on the initial state of the central
	qubit as before, namely , $|a_0(0)|=|a_1(0)|=1/\sqrt{2}$.
	In terms of phase factors, the fulfillment of the equality condition
	requires all phase factors in each group to align 
	$\phi_{n_{I/II}}(t)-\phi_{m_{I/II}}(t)=2j\pi$, for some integer $j$,
	while the fulfillment of the orthogonality condition requires strong misalignment
	of the phases from different groups with respect to each other
	$\phi_{n_{I}}(t)-\phi_{n_{II}}(t)=(2m+1)\pi$.
	These conditions are equivalent to the phase condition for a qubit environment
	(the equality condition was automatically fulfilled in that case,
	since there was only one state in each group), but obviously the number of phase
	conditions that have to be simultaneously met grows strongly with
	the size of the environment. Hence, it is harder to obtain the situation
	when all conditions are met for a larger environment, even though there are no
	additional constraints on the division of environmental states between the two 
	groups (as long as two groups exist).
	
	To see that such glimpses of objectivity at discrete times are possible even
	for larger environments, it is again convenient to consider the asymmetric coupling
	case, similarly as in case of the environment size restricted to one qubit.
	Then the different phases are linear functions of time and only the differences
	between the eigenvalues of $\hat{V}^1$, $v^1_n$, are relevant. Obviously,
	if $\hat{V}^1$ is degenerate and has only two different eigenvalues, the situation
	operationally reduces to the qubit environment case (and the states are permanently
	divided into the two groups by these different values). Otherwise, at different 
	instances of time different subgroups of environmental states can align,
	leading to a more complicated pattern of SBS state emergence 
	(typically, for randomly chosen $v^1_n$, SBS states will be glimpsed more 
	sparsely throughout the evolution with growing size of the environment).
	In time instances that the phase factors corresponding to the effect of the environment
	align into two groups, leading to opposite phase factors,
	the SBS state will emerge, and will be of the form (\ref{chuj}),
	where the central qubit states $|\psi_{I/II}\rangle$
	can still only be a trivial variation of the states given in eqs (\ref{pm}).

	\subsection{Beyond the qubit \label{sec4b}}
	
	The fulfillment of all separability criteria, (\ref{sep}) and (\ref{sep2}), allows to transform
	the full system-environment density matrix into an obviously
	separable form as in the case of the qubit.
	For a single environment it looks exactly the same as eq.~(\ref{pot2}),
	but the qubit states are now replaced by states of 
	the system dimension,
	\begin{equation}
	\label{psi2}
	|\psi_n\rangle=\sum_{i=0}^{d_Q-1}a_i(t)e^{-i\phi_{n_k}^i(t)}|i\rangle,
	\end{equation}
	where $|i\rangle$ labels the pointer states of the system as before,
	$a_i(0)$ are the initial occupations of the system state
	and the coefficients $a_i(t)$ contain the free evolution of the system
	and $d_Q$ denotes the dimension of the system.
	
	The state (\ref{pot2}) still has zero discord with respect to 
	the environment and the environmental states $|n(t)\rangle$
	still form an orthonormal basis. Therefore what is necessary for
	the emergence of SBS states is that the qubit $|\psi_n\rangle$ states
	would form an orthonormal basis. For this to occur the initial central system
	occupations must all be equal, $|a_i(0)|=1/\sqrt{d_Q}$ (since all
	states $|\psi_n\rangle$ have the same occupations)
	while the different phase factors in eq.~(\ref{psi2})
	must align so that the $|\psi_n\rangle$
	states form a basis. As a result, they form a MUB with respect to the $|i\rangle$ base \cite{durt04,brierley10}.
	For systems
	which can be decomposed into qubits, 
	so their dimension is $d_Q=2^m$, with integer $m$,
	such bases can be easily constructed by tensor multiplication of the qubit equal-superposition basis.
	Note that for glimpses of objectivity to be possible, the dimension of the 
	environment must be of the same size or larger than that of the central system.
	
    For a central system state larger than a qubit
	(plus environment) to be a true
	SBS state, all $d_Q$ central system MUB states must be present in the decomposition
	(\ref{pot2}), one should therefore expect the natural occurrence of
	SBS states to be less frequent also with the growing 
	size of the central system, unless the environment is specially structured, so the phase relations
	between the different pointer states in the states $|\psi_n\rangle$
	oscillate with frequencies which are not random.
	
	\section{Multiple environments \label{sec5}}
	
Let us now consider more environments than just one, while again restricting
the central system to a qubit. Assuming the initial state of the environment is a product, $\hat \rho^E(0)=\bigotimes_k\hat \rho^k(0)$, we obtain a very similar form 
of the time evolved qubit-environments density matrix to
the single-environment density matrix (\ref{pot2}),
\begin{eqnarray}
		\label{pot3}
		\hat{\sigma}(t)=&&\sum_{n_1,\dots,n_N}\left(\prod_{k=1}^N p_{n_k}\right)
		|\psi_{n_1\dots n_N}(t)\rangle\langle\psi_{n_1\dots n_N}(t)|\nonumber\\
		&& \bigotimes_{q=1}^N |n_q(t)\rangle\langle n_q(t)|,
		\end{eqnarray}
		with the states of the different environments now resolved
		and their evolved states are obtained from the initial eigenstates
		of each density matrix (\ref{ntilde}) as previously, $|n_q(t)\rangle=\hat{w}_i^q(t)|\tilde{n}_q\rangle$,
		where $\hat{w}_i^q(t)$ is given by eq.~(\ref{wi}).
		The qubit states also retain a form similar to the single-environment
		case, with the exception that the phases now accumulate from different
		environments,
		\begin{equation}
		\label{stan2}
		|\psi_{n_1\dots n_N}(t)\rangle \equiv a|0\rangle+b(t)e^{-i\sum_k\phi_{n_k}(t)}|1\rangle .
		\end{equation}
		
		For the qubit-multiple-environments density matrix (\ref{pot3}) to be of SBS
		form (\ref{sbs}), firstly the conditions for the orthogonality qubit states 
		$|\psi_{n_1\dots n_N}(t)\rangle$ should be fulfilled. These conditions
		have been specified in the previous section, and the ones which apply here
		are those pertaining to a single larger environment. Namely each qubit state
		$|\psi_{n_1\dots n_N}(t)\rangle$
		must at the given time $t$, when objectivity is to be glimpsed, be one of two orthogonal
		states as in Sec.~\ref{sec4a}.
		The only difference is that the phases accumulated by each qubit state
		are now a sum of the phases stemming from the evolution of each environment.
		
		If the SBS conditions for the qubit part of the density matrix are met,
		there is still the question if the SBS form is retained on the side
		of each environment.
		For a single environment this was always the case, since at any time $t$
		the environmental states $|n(t)\rangle$ form an orthonormal basis, so states
		corresponding to either group of qubit states can be combined into a diagonal
		density matrix in this basis, and the two density matrices must be
		orthogonal to one another.
		
		In case of multiple environments, this simple method does not work, because
		the states $\bigotimes_{q=1}^N |n_q(t)\rangle$, although they are composed of parts
		which are orthogonal 
		for every environment, will in general loose that property when grouped as in eq.~(\ref{rhoI_II}).  
		To see this, let us consider two environments composed of a single qubit
		each and denote the environmental states $|n_1(t)\rangle\otimes|n_2(t)\rangle$
		corresponding to a time $t$ when the central qubit fulfills the SBS
		conditions as $|00\rangle$, $|01\rangle$, $|10\rangle$, and $|11\rangle$.
		Let us further denote the two states of the central qubit as $|+\rangle$
		and $|-\rangle$. In the symmetric case, when two of the environmental states
		correspond to central qubit state $|+\rangle$, say
		$|00\rangle$ and $|01\rangle$, and the other two to the $|-\rangle$ state,
		the states of the central qubit will be fully distinguishable by measurements
		on one qubit, but completely indistinguishable by measurements on the other.
		For the allotment as chosen above, the density matrices of each environment 
		that enter eq.~(\ref{sbs}) are given by $\hat{\rho}_0^1=|0\rangle\langle 0|$,
		$\hat{\rho}_1^1=|1\rangle\langle 1|$, and 
		$\hat{\rho}_0^2, \hat{\rho}_1^2 \sim p|0\rangle\langle 0|+
		(1-p)|1\rangle\langle 1|$ so the states of the second environment are clearly not orthogonal to each other.
		In the asymmetric case, when only one state is allotted to state $|+\rangle$, say
		$|00\rangle$, and the other three to state $|-\rangle$, the qubit-environments
		density matrix can no longer be written in the SBS form (\ref{sbs}) corresponding
		to two environments, since we have
		\begin{eqnarray}
		\hat{\sigma}_{SBS}&=&p_+ |+\rangle\langle +|\otimes|0\rangle\langle 0|
		\otimes|0\rangle\langle 0|\\
		\nonumber
		&&
		+p_-^A |-\rangle\langle -|\otimes|0\rangle\langle 0|
		\otimes|1\rangle\langle 1|\\
		\nonumber
		&&
		+\frac{p_-^B}{2}|-\rangle\langle -|\otimes|1\rangle\langle 1|
		\otimes\left(|0\rangle\langle 0|+
		|1\rangle\langle 1|\right).
		\end{eqnarray}
		Here, the measurement $|0\rangle$ or $|1\rangle$ on either environmental qubit
		can correspond to either central qubit state.
		
		The situation only becomes more complicated for larger central systems,
		but the outcome is the same and SBS states obtained in the described way
		can occur in nonentangling evolutions only if there is one observed environment.

	\section{Conclusion \label{sec6}}
	
	We have studied a class of nonentangling evolutions
	which lead to pure dephasing of the central system due to the interaction
	with one or multiple environments.
	We have shown that system-environment entanglement is not
	necessary for the emergence of SBS states, but the situation
	when it does emerge
	is rather special. Firstly, it does not require there to be unobserved environments
	(no decoherence source is necessary).
	Secondly, it is possible only at discrete instants of time, and thirdly,
	it can only occur  if the initial state of this system is an equal-superposition state (so only in  mutually unbiased bases of the central system). 
	Furthremore, it is only possible for one observed environment.
	
	If the central system is a qubit initially in an equal 
	superposition state,
	SBS states will occur naturally throughout the evolution
	at instances of time when the phases which are the outcome of the 
	difference of the evolution of the environment conditional on 
	the state of the qubit
	all reach one of two values which
	differ by $\pi$ (states corresponding to both types of phases must exist)
	yielding orthogonal qubit states for the qubit. 
	Contrarily to the standard way of obtaining SBS states, orthogonality of 
	environmental states is always fulfilled in the studied scenario, while
	the orthogonality of the qubit states is problematic
	(it is guaranteed when the qubit basis is its pointer basis 
	as in the standard scenario), and can occur only
	at discrete points of time. We have shown that 
	such instances must occur for asymmetric qubit-environment couplings,
	for which the aforementioned phases are linear functions of time. 
	The glimpses of objectivity will occur for environments larger than a qubit, 
	but since this requires the alignment of the number of phase factors
	equal to the dimension of the environment, they will occur 
		the more rarely the bigger the environment (assuming that
		the different states of the environment oscillate differently).

\end{document}